\documentclass[aps,pre,twocolumn,groupedaddress]{revtex4-1}

\usepackage{subfigure}
\usepackage{graphicx}
\usepackage{bm}
\usepackage{amsmath}
\usepackage{epstopdf}
\usepackage{dcolumn}

\newcommand{\STO}{SrTiO$_3$}
\newcommand{\LAO}{LaAlO$_3$}
\newcommand{\Tc}{$T_c$}
\newcommand{\Vg}{$V_g$}
\newcommand{\Hcpar}{$H_{c\parallel}$}
\newcommand{\Hcperp}{$H_{c\perp}$}

\newcommand{\etal}{\emph{et al.}}

\newcommand{\Rs}{$R_S$}

\begin{document}

\title{Link between the Superconducting Dome and Spin-Orbit Interaction in the (111) \LAO/\STO~ Interface}
\author{P. K. Rout}
\affiliation{Raymond and Beverly Sackler School of Physics and Astronomy, Tel-Aviv University, Tel Aviv, 69978, Israel}
\author{E. Maniv}
\affiliation{Raymond and Beverly Sackler School of Physics and Astronomy, Tel-Aviv University, Tel Aviv, 69978, Israel}
\author{Y. Dagan}
\affiliation{Raymond and Beverly Sackler School of Physics and Astronomy, Tel-Aviv University, Tel Aviv, 69978, Israel}
\email[]{yodagan@post.tau.ac.il}

\date{\today}

\begin{abstract}
We measure the gate voltage (\Vg) dependence of the superconducting properties and the spin-orbit interaction in the (111)-oriented \LAO/\STO~ interface. Superconductivity is observed in a dome-shaped region in the carrier density-temperature phase diagram with the maxima of superconducting transition temperature \Tc~and the upper critical fields lying at the same \Vg. The spin-orbit interaction determined from the superconducting parameters and confirmed by weak-antilocalization measurements follows the same gate voltage dependence as \Tc. The correlation between the superconductivity and spin-orbit interaction as well as the enhancement of the parallel upper critical field, well beyond the Chandrasekhar-Clogston limit suggest that superconductivity and the spin-orbit interaction are linked in a nontrivial fashion. We propose possible scenarios to explain this unconventional behavior.
\end{abstract}

\maketitle

Oxide heterostructures provide unique platform where various degrees of freedom from the constituent materials can combine such that new collective phenomena emerge at the interfaces \cite{hwang2012emergent}. An interesting example is a two-dimensional (2D) electron liquid at the interface between (100)-oriented \STO~ and \LAO~ that exhibits gate tunable superconductivity \cite{caviglia2008electric,maniv2015strong,shalom2010tuning} and spin-orbit interaction \cite{shalom2010tuning, caviglia2010tunable,liang2015nonmonotonically}. Recent experiments on (111) \LAO/\STO~ have shown 2D conduction \cite{herranz2012high,davis2017anisotropic,rout2017six} and superconductivity with a transition temperature (\Tc) of about 100 mK \cite{monteiro2017two, davis2017superconductivity}. In a (111)-oriented \LAO/\STO~ interface, the cubic lattice is projected onto the (111) plane of the interface, resulting in a 2D sixfold crystalline structure. Angle-resolved photoemission studies on the (111) \STO~ surface reveal a sixfold symmetric electronic structure \cite{walker2014control, rodel2014orientational}. This 2D crystalline symmetry is also reflected in the magnetotransport properties \cite{rout2017six} and has been predicted to host exotic electronic orders \cite{xiao2011interface, doennig2013massive,scheurer2017selection,okamoto2017transition}. At low temperatures, this symmetry is lowered, since bulk \STO~ undergoes multiple structural transitions. Below 105 K, a transition from a cubic to a tetragonal phase occurs \cite{muller1979srti}. The symmetry is further reduced to triclinic below $\sim$70 K, and polar domain walls where inversion symmetry is broken are created \cite{salje2013domains}. Such a domain wall can be pinned to the interface, resulting in unconventional superconductivity, which is linked to spin-orbit coupling.
\par
In a 2D superconductor, for a magnetic field applied perpendicular to the superconducting plane, superconductivity is broken when vortices become closely packed. By contrast, the parallel upper critical field (\Hcpar) is determined by the Chandrasekhar-Clogston limit \cite{chandrasekhar1962note,clogston1962upper}, which is set by comparing the Zeeman energy to the superconducting gap. In the presence of a spin-orbit interaction, this upper bound is relaxed \cite{klemm1975theory,nakamura2013multi}.
\par
In this Letter, we report a nonmonotonic (dome-shaped) dependence of \Tc~ with a gate voltage in the (111) \STO/\LAO~ interfaces. From the gate dependence of \Tc~ and \Hcpar~, we estimate the spin-orbit energy ($\varepsilon_{SO}$), which follows the nonmonotonic behavior of \Tc. Remarkably, we found similar behavior for the spin-orbit field  $H_{SO}$ extracted from weak antilocalization measurements.
\par
Epitaxial films of LaAlO$_3$ were deposited on an atomically flat SrTiO$_3$ (111) substrate using pulsed laser deposition. The details of the deposition procedure and substrate treatment are described in Ref. \cite{rout2017six}. We control the layer-by-layer growth of 14 monolayers (LaO$_3$/Al layers) by reflection high-energy electron diffraction oscillations. The atomic force microscope images show the step and terrace morphology of the film with step heights of 0.22 nm. The electrical measurements with the current along the [11$\bar{2}$]  direction were carried out in a Leiden Cryogenics custom-made dilution refrigerator.

\begin{figure}
\begin{center}
\includegraphics[width=1\hsize]{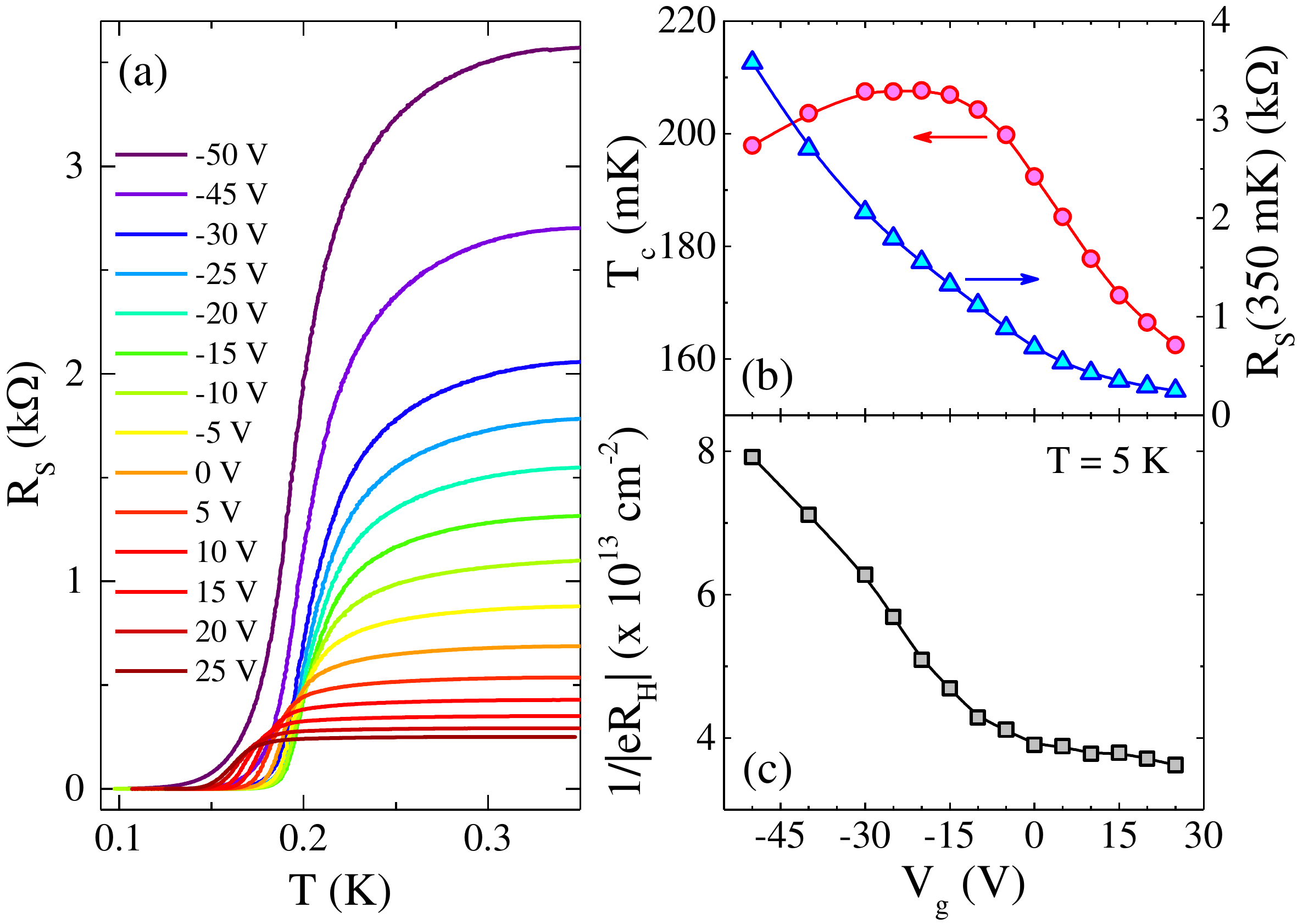}
\caption{(a) Temperature dependence of sheet resistance \Rs~($T$) for various gate voltages. (b) \Tc~ and \Rs~(350 mK) as a function of \Vg. (c) Gate dependence of the inverse Hall coefficient 1/$\left| {eR_H} \right|$ at $T =$5 K.}
\label{Fig-SC}
\end{center}
\end{figure}

Figure~\ref{Fig-SC} (a) presents the temperature-dependent sheet resistance $R_S$ ($T$) at various gate voltages \Vg. A clear gate-dependent superconducting transition is observed. We define the critical temperature \Tc~ as the temperature at which $R_S$ reaches half of its value at 350~mK. The normal state resistance $R_S$ (350 mK) decreases monotonically with increasing \Vg~ [Fig.~\ref{Fig-SC} (b)], which is consistent with previous reports \cite{davis2017anisotropic,rout2017six}. The monotonic increase of $R_S$ is contrasted with the nonmonotonic dependence of \Tc~ on \Vg. A similar dome-shaped region in the carrier density-temperature phase diagram is seen in many unconventional superconductors and in the (100) \LAO/\STO~ interface.
\par
In the (100) \LAO/\STO~ interface, the Hall coefficient depends nonmonotonically on the gate voltage. Surprisingly, this nonmonotonic behavior is also seen in the gate dependence of the Shubnikov–de Haas oscillations (SdH) frequency. Both the SdH frequency and low field inverse Hall coefficient follow the gate dependence of \Tc~ for the (100) interface \cite{maniv2015strong,smink2017gate}, or the superconductivity starts appearing when the low field inverse Hall coefficient decreases from its maximum value \cite{singh2017competition}. By contrast, for the (111) interface the inverse Hall coefficient monotonically decreases with \Vg~ [Fig.~\ref{Fig-SC} (c)] consistent with previous observations \cite{davis2017anisotropic, rout2017six}. In the case of the (111) \LAO/\STO~ interface, the titanium t$_{2g}$ bands are split into low and high spin states due to the atomic spin-orbit interaction \cite{xiao2011interface, doennig2013massive}. We have shown that the lower spin state is first populated when accumulating electrons with increasing \Vg~\cite{rout2017six}. This two-band scenario complicates the interpretation of the Hall data. We have estimated the amount of carrier density modulation due to the electric field effect similar to Refs. \cite{caviglia2008electric,biscaras2012two}. Since the \Vg~range used is relatively small, the nonlinearities in the dielectric constant ($\epsilon$) can be neglected and thus the corresponding modulation of electron density is $\simeq$ 1.3 $\times$ 10$^{13}$ cm$^{-2}$ with $\epsilon \simeq$ 15000. This value is much smaller than the net change in 1/$\left| {eR_H} \right|$ of $\simeq$ 4.3 $\times$ 10$^{13}$ cm$^{-2}$. Moreover, the electron density due to the field effect increases with \Vg~ in contrast to the observed behavior in Fig.~\ref{Fig-SC} (c). All these observations indicate the presence of a hole band in addition to electron band(s) in the (111) interface. We have confirmed this scenario by analyzing the normal state transport data via a simplistic noninteracting two-band model with one hole and one electron band (see Ref. \cite{Supple} for more details). Therefore,  it is possible that the hole contribution to the electronic transport (and perhaps to superconductivity) becomes important in this \Vg~range \cite{davis2017anisotropic}. This is also consistent with the polar structure of the (111) interface \cite{herranz2012high}.

\begin{figure*}
\begin{center}
\includegraphics[width=0.8\hsize]{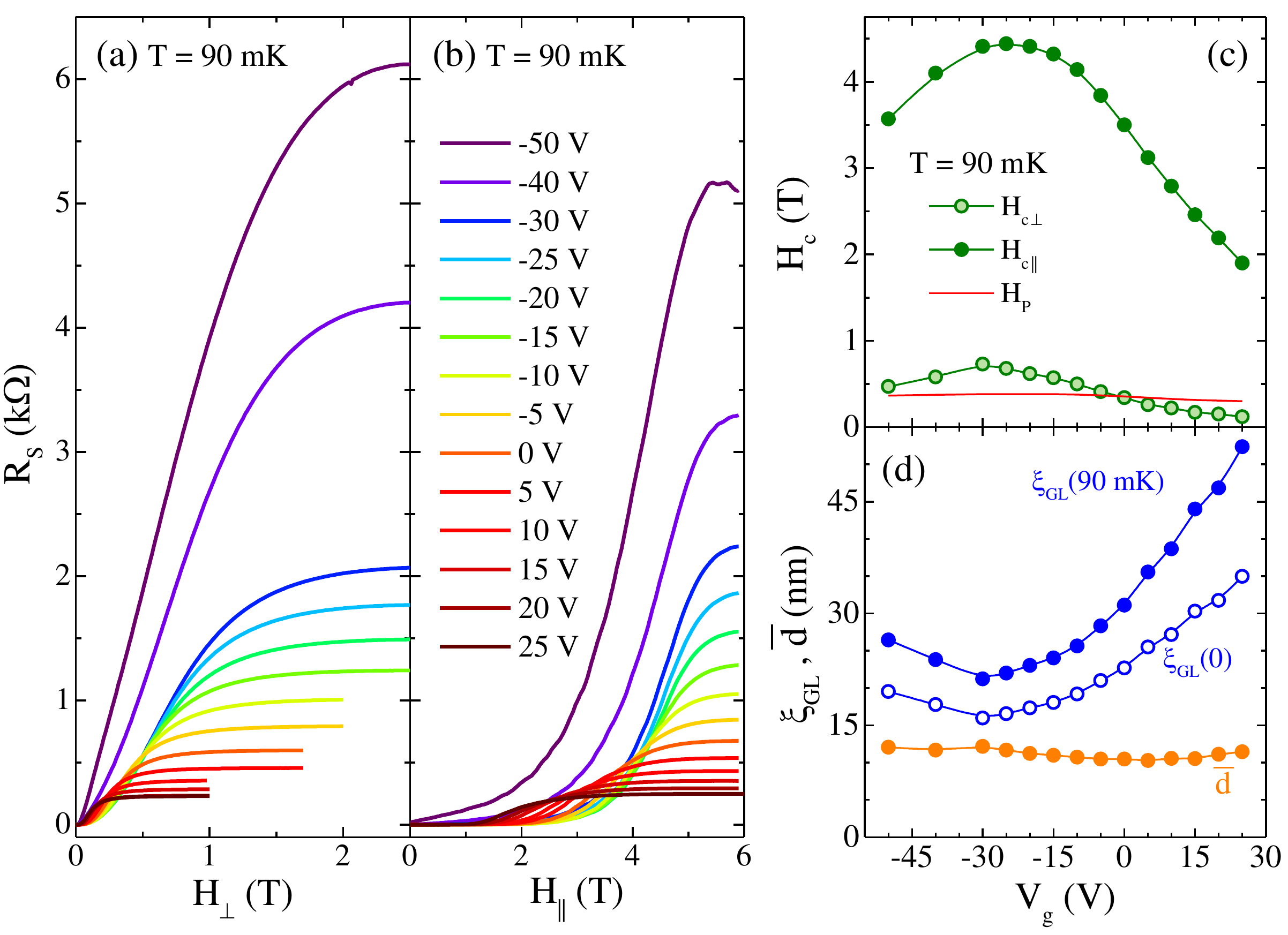}
\caption{ Magnetoresistance \Rs~($H$) at $T =$ 90 mK in (a) perpendicular ($\vec{H}$ perpendicular to the current and interface) and (b) longitudinal ($\vec{H}$ parallel to the current and interface) configurations for various \Vg. (c) \Hcperp~ and \Hcpar~ at 90 mK as a function of \Vg~ along with the Chandrasekhar-Clogston limit $H_P$. (d) Gate dependence of $\xi_{GL}$ (90 mK), $\xi_{GL}$(0), and $\overline{d}$.}
\label{Fig-Hc}
\end{center}
\end{figure*}
\par

The sheet resistance versus magnetic field at 90~mK for various gate voltages is plotted in Figs.~\ref{Fig-Hc} (a) and~\ref{Fig-Hc} (b) for perpendicular and parallel field configurations, where the sample is properly aligned with the field within an accuracy of 2$^{\circ}$. We define the critical field (\Hcperp) for the perpendicular magnetic field configuration such that $R_S$(\Hcperp)$=R_S$ (350~mK)/2 and a similar criterion is followed for \Hcpar \cite{Remark}. In Fig.~\ref{Fig-Hc} (c) we plot \Hcpar~ and \Hcperp~ as a function of \Vg~ both exhibiting nonmonotonic behavior with the maximum at the same gate voltage as \Tc. \Hcpar$>$\Hcperp~ for all gate voltages reaching a maximal ratio of $\sim$16. Such strong anisotropy between two field orientations is evidence for 2D superconductivity in the (111) interface. Thus, it is expected that the superconducting layer thickness ($d$) should be smaller than the Ginzburg-Landau coherence length ($\xi_{GL}$). To check this, we extract $\xi_{GL}$ from \Hcperp~ using the relation: ${\xi _{GL}} = \sqrt {{\Phi _0}/2\pi {H_{c \bot }}} $ . It is presented in Fig.~\ref{Fig-Hc} (d) together with its extrapolation to zero temperature using \Hcperp($T$)=\Hcperp(0)$(1-T/T_c)$ valid for a 2D superconductor. Since the parallel magnetic field fully penetrates a 2D $(d \ll \xi)$ superconductor we can only estimate the upper limit for $d$ denoted as $\overline{d}$, which can be found from $\overline{d} = \sqrt 3 {\Phi _0}/\pi {\xi _{GL}}{H_{c\parallel }}$ [see Fig.~\ref{Fig-Hc} (d)]. We note that, for all \Vg, $\overline{d}<\xi_{GL}(0)$, rendering superconductivity in the (111) \STO/\LAO~ two dimensional.
\par
For a parallel field configuration in a 2D superconductor, the orbital motion and vortices can be neglected making the Zeeman energy the dominant pair-breaking effect. This leads to an upper (Chandrasekhar-Clogston) limit of \Hcpar~ given by $H_P = 3.5k_BT_c/\sqrt{2}g\mu_B$ ($\mu_B$ is the Bohr magneton) in the BCS weak coupling limit \cite{chandrasekhar1962note,clogston1962upper}. Assuming a gyromagnetic ratio of $g\simeq$2, we observe \Hcpar$>H_P$ for all gate voltages reaching a maximal ratio of $\sim$11 [Fig.~\ref{Fig-Hc} (c)]. In the presence of strong spin-orbit coupling the Chandrasekhar-Clogston limit can be relaxed. Other reasons for breaking this limit could be strong coupling superconductivity, many-body effects, and an anisotropic pairing mechanism.

\par
\begin{figure*}
\begin{center}
\includegraphics[width=1\hsize]{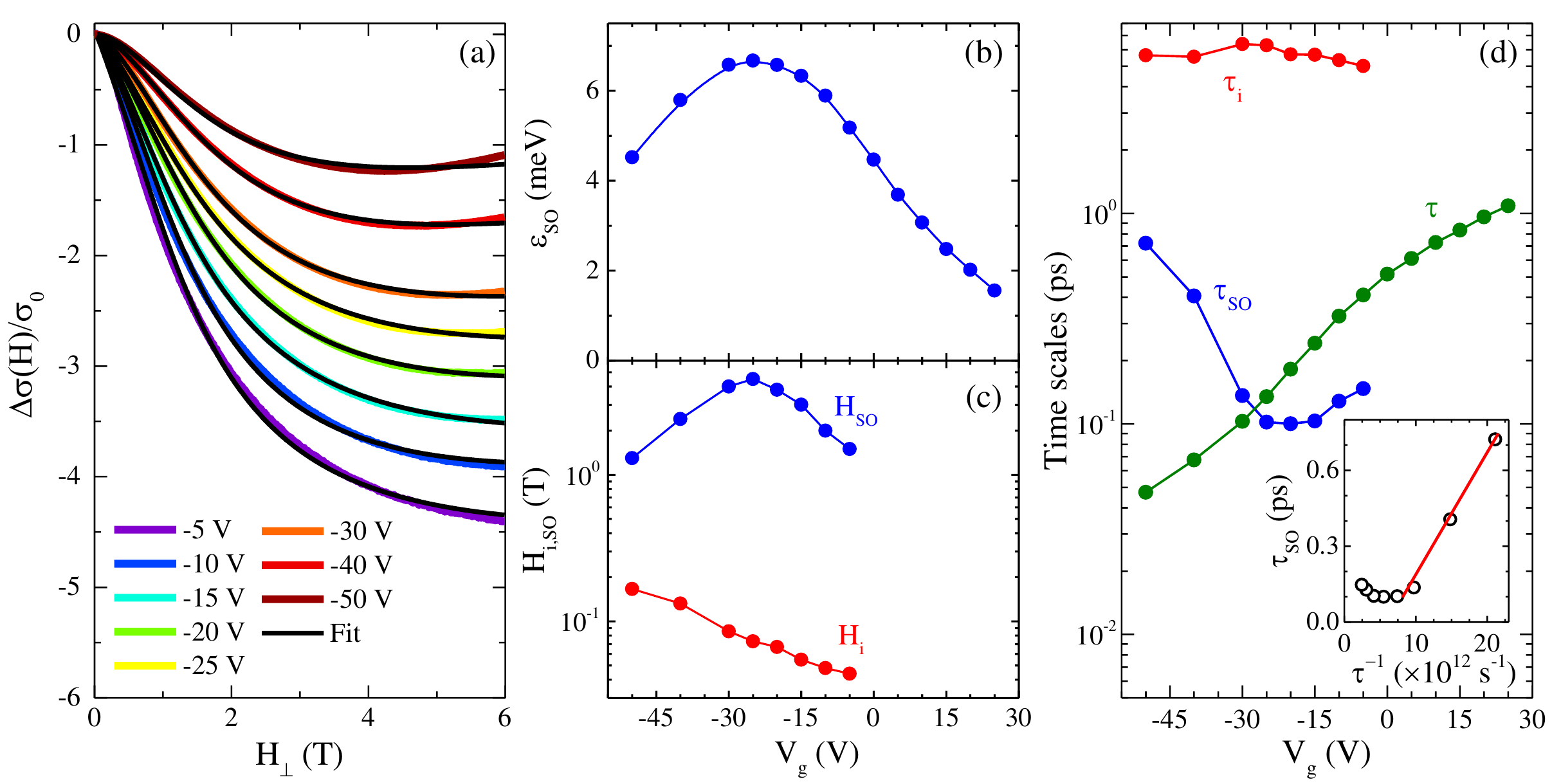}
\caption{(a) The normalized perpendicular magnetoconductance $\Delta \sigma (H)/\sigma _0$ for different \Vg~ at $T =$ 1.3 K. The black solid lines are the fits according to Eq. \ref{WAL}. (b) $\varepsilon_{SO}$ as  a function of \Vg~ determined from \Hcpar~ (see the text for more details). (c) Gate dependence of ${H_i }$ and ${H_{SO} }$ extracted from the fitting of weak antilocalization. (d)  Gate dependence of $\tau_{i}$, $\tau _{SO}$, and $\tau$. The inset shows $\tau _{SO}$ as a function of $\tau^{-1}$ along with the solid line as a guide to the eye.}
\label{Fig-WAL}
\end{center}
\end{figure*}

To determine the spin-orbit interaction from \Hcpar~, we use a somewhat oversimplified picture of spin-orbit scattering that suppresses spin orientation by the Zeeman field \cite{klemm1975theory}. For a strong spin-orbit interaction, \Hcpar~ can be expressed in terms of the spin-orbit energy ($\varepsilon_{SO}$) as ${H_{c\parallel}} = 0.602\sqrt { \varepsilon_{SO} /{k_B}{T_c}} {H_P}$ with $\varepsilon_{SO}= \hbar/\tau _{SO}$, and $\tau _{SO}$ is the spin-orbit scattering time. Remarkably, this analysis reveals a nonmonotonic dependence of $\varepsilon_{SO}$ on \Vg~ as shown in Fig.~\ref{Fig-WAL} (b). This is the main finding of our Letter. For (110) \LAO/\STO~, gate-independent spin-orbit coupling has been observed \cite{herranz2015engineering}; perhaps because of the nonpolar structure of this interface. The findings on the (110) interface are contrasted with our results of a strong and gate-tunable spin-orbit interaction for the (111) interface that follows the behavior of the superconducting dome. A weaker correlation between spin-orbit coupling and \Tc~ in the (100) interface can be deduced by combining Refs. \cite{shalom2010tuning, caviglia2010tunable,liang2015nonmonotonically}, where \Hcpar is smaller.

\par

To further confirm the presence of a spin-orbit interaction, we studied the perpendicular magnetoresistance well above \Tc~ at 1.3 K [Fig.~\ref{Fig-WAL} (a)]. For a 2D diffusive metallic system placed in a perpendicular magnetic field ($H$), the field-dependent quantum correction to conductivity $\Delta \sigma (H)$ normalized by quantum conductance ($\sigma _0=2e^2/h$) can be expressed as \cite{maekawa1981magnetoresistance,caviglia2010tunable}
\begin{equation}\label{WAL}
\begin{split}
\frac{{\Delta \sigma (H)}}{{{\sigma _0}}} = \Psi \left( {\frac{H}{{{H_i} + {H_{SO}}}}} \right)\\
+ \frac{1}{{2\sqrt {1 - {\gamma ^2}} }} \Psi \left( {\frac{H}{{{H_i} + {H_{SO}}(1 + \sqrt {1 - {\gamma ^2}} )}}} \right)\\
- \frac{1}{{2\sqrt {1 - {\gamma ^2}} }} \Psi \left( {\frac{H}{{{H_i} + {H_{SO}}(1 - \sqrt {1 - {\gamma ^2}} )}}} \right)\\
 - \frac{{A{H^2}}}{{1 + C{H^2}}}.
\end{split}
\end{equation}
where $\Psi (x) = \ln (x) + \psi \left( {\frac{1}{2} + \frac{1}{x}} \right)$ [$\psi (x)$  is the digamma function] and $\gamma  = g{\mu _B}H/4eD{H_{SO}}$ ($D$ is the diffusion coefficient). $H_{i}$ and $H_{SO} $ are the inelastic and spin-orbit fields, respectively. The classical orbital magnetoresistance contributes a Kohler term to Eq.~(\ref{WAL}) with the parameters $A$ and $C$. Figure~\ref{Fig-WAL} (c) shows $H_{i}$ and $H_{SO}$ for different \Vg~ (see Supplemental Material for the gate dependence of $g$, $A$, and $C$ \cite{Supple}). Clearly, $H_{SO}  > H_{i}$ for all \Vg~, suggesting that we are in the weak antilocalization regime [see Fig.~\ref{Fig-WAL} (a)]. $H_{SO}$ from weak antilocalization [Fig.~\ref{Fig-WAL} (c)] shows nonmonotonic behavior similar to $\varepsilon_{SO}$ inferred from superconductivity [Fig.~\ref{Fig-WAL} (b)], and, furthermore, they have maximum value at the same gate voltage as \Tc.
\par
In general, the \LAO/\STO~ interface has a complicated band structure involving multiple contributions from the titanium $d$ bands \cite{shalom2010shubnikov, lerer2011low}. Therefore, the extracted parameters from weak antilocalization do not correspond to an individual band; instead an averaged value over all the bands should be considered \cite{rainer1985multiband}. We have extracted various averaged time scales, i. e. $\tau _{SO}$, $\tau _{i}$ (inelastic time), and $\tau$ (elastic scattering time) [Fig.~\ref{Fig-WAL} (d)]. The $\tau _{SO (i)}$ are related to $H_{SO (i)}$ determined  from weak antilocalization as ${H_{SO (i)}} = \hbar /4eD{\tau _{SO (i)}}$. The effective diffusion coefficient ($D$) and $\tau$ are calculated using a na\"{\i}ve Drude model for a 2D electron gas (see Ref. \cite{Supple}). Using this analysis we find that $\tau _{SO}$ depends linearly on $\tau^{-1}$ for \Vg~ $<$ -25 V [see the inset in Fig.~\ref{Fig-WAL} (d)] while for \Vg~ $>$ -25 V both $\tau_{SO}$ and $\tau$ increase with \Vg~[Fig.~\ref{Fig-WAL} (d)].
\par
The low \Vg~regime (\Vg~$<$ -25 V) is governed by a D'yakonov-Perel'-type spin-orbit relaxation mechanism for which $\tau_{SO}\propto \tau^{-1}$. In this scenario the electron precesses around the spin-orbit field, which is changing due to momentum scattering at a typical time $\tau$ \cite{vzutic2004spintronics}. The high \Vg~ regime, on the other hand, is characterized by $\tau_{SO}\propto \tau$, suggesting that the electron spin is coupled to the crystal momentum. Interestingly these two regimes separated by the point where $\tau_{SO} \simeq \tau$ and the maximum of \Tc~(and \Hcpar) dome lies close to this \Vg. All these observations suggest the mixing of multiple bands in the presence of a strong spin-orbit interaction for higher \Vg. This scenario concurs with our recent report of crystalline sixfold anisotropic magnetoresistance in the (111) interfaces \cite{rout2017six}, where the sixfold term appears as a result of another band with higher spin state $J$ getting populated with increasing \Vg. It is therefore possible that the crystalline spin-orbit interaction becomes important close to this avoided band crossing region due to the orbital mixing \cite{zhong2013theory,nakamura2013multi}. This interaction becomes smaller as \Vg~is further increased away from the band crossing regime, resulting in a dome in the spin-orbit energy versus \Vg. Such a multiband effect can also lead to dome-shaped superconductivity with maximum \Tc~lying at this regime [as observed in Fig.~\ref{Fig-SC} (b)] similar to the case for the (100) interface \cite{maniv2015strong}. A more exotic mechanism of superconductivity in the \LAO/\STO~interface involves the formation of a Fulde-Ferrell-Larkin-Ovchinikov (FFLO) state due to large spin-orbit coupling  \cite{michaeli2012superconducting}. This can somewhat explain the nonmonotonic gate dependence of \Hcpar~and \Tc~with the maxima lying at $\tau_{SO} = \tau$. However, the \Hcpar~for a quasi-2D superconductor in a FFLO state is estimated to be at most 2.5 times the Chandrasekhar-Clogston limit \cite{shimahara1997fulde}, which is much lower than the observed values [see Fig.~\ref{Fig-Hc} (c)]. Therefore, a full theoretical understanding of the phenomenological link observed here between the superconducting dome and the spin-orbit energy is yet to be developed.
\par
Salje \etal~have found that for \STO~ below $\sim$70 K the tetragonal symmetry is lowered and the Sr atoms are displaced along the [111] direction leading to the breaking of local inversion symmetry \cite{salje2013domains}. It is therefore possible that a (111) \STO-based polar interface has such broken inversion symmetry in addition to conventional inversion symmetry breaking observed at polar oxide interfaces, which can result in an unconventional superconductivity. It has been recently suggested that dichalcogenide monolayers with hexagonal structure can be a realization of exotic Ising superconductivity where the spins are locked in an out-of-plane configuration due to the breaking of centrosymmetry \cite{lu2015evidence,xi2016ising,saito2016superconductivity}.  We also note that the possibility for a nodeless time-reversal-symmetry-breaking superconducting order parameter has been proposed for (111) \STO-based interfaces from symmetry considerations \cite{scheurer2017selection}.
\par
In summary, the superconducting transition temperature \Tc~ of the (111) \LAO/\STO~ interface has a nonmonotonic dependence on the gate voltage. Maximum \Tc~ is found at the same gate voltage where maximal values of spin-orbit field H$_{SO}$ and spin-orbit energy $\varepsilon_{SO}$ are observed. H$_{SO}$ is extracted from weak antilocalization while $\varepsilon_{SO}$ is estimated from the superconducting properties. The \Hcpar~ exceeds the Chandrasekhar-Clogston limit by more than an order of magnitude due to a strong spin-orbit interaction. We suggest that the crystalline spin-orbit interaction becomes important close to an avoided band crossing region. In this regime orbital mixing can lead to enhanced spin-orbit interaction and superconductivity, which become weaker as \Vg~ is tuned away from this avoided band crossing regime. This results in a dome in the spin-orbit energy (and \Tc~) versus \Vg. However, a deeper insight to the link between spin-orbit interaction and the superconducting dome requires further development of theoretical models for this unique hexagonal oxide interface.
\par

P.K.R. and E.M. contributed equally to this work. We are indebted to Moshe Goldstein and Alexander Palevski for useful discussions. This work has been supported by the Israel Science Foundation under Grant No. 382/17, the Israel Ministry of Science technology and space under Contract No. 3-11875 and the Bi-national science foundation under Grant No. 2014047.

\clearpage

\newcommand{\beginsupplement}{%
        \setcounter{section}{0}
        \renewcommand{\thesection}{S\arabic{section}}%
        \setcounter{table}{0}
        \renewcommand{\thetable}{S\arabic{table}}%
        \setcounter{figure}{0}
        \renewcommand{\thefigure}{S\arabic{figure}}%
        \setcounter{equation}{0}
        \renewcommand{\theequation}{S\arabic{equation}}%
     }

\begin{widetext}
\beginsupplement

\section*{\textbf{Supplemental Material}}

\section{Two-band analysis of Hall data}
We have analysed the Hall data (Fig. \ref{Fig-HallTwoBand}) using a simplified two band model with no interaction effects. As discussed in the manuscript, one of them should be a hole band while the other is a electron band. In this model, the low field Hall coefficient is given as:
\begin{equation}\label{Eq. S1}
{R_H} = \frac{1}{e}\frac{{{n_h}\mu _h^2 - {n_e}\mu _e^2}}{{{{({n_h}{\mu _h} + {n_e}{\mu _e})}^2}}}.
\end{equation}
where $n_{h(e)}$ and $\mu_{h(e)}$ are hole (electron) carrier density and mobility, respectively. The corresponding sheet resistance at zero magnetic field is $R_S = [e({n_h}{\mu _h} + {n_e}{\mu _e})]^{ - 1}$. For relatively small \Vg~range used,  the nonlinearities in the dielectric constant can be neglected and the change in total carrier density for change in \Vg~($\Delta V_g$) can be given as: ${C_{STO}}\Delta {V_g} = e(\Delta {n_h} + \Delta {n_e})$, where ${C_{STO}}$ is the capacitance of STO (111) per unit area. The corresponding changes in $n_{h(e)}$ are proportional to effective masses $m_{h(e)}$ for parabolic bands and therefore $\Delta {n_h}  / \Delta {n_e} = m_{h}/m_{e}$. We have assumed $m_{h} = m_{e}$ for our calculations. Using all these expressions, we have extracted $n_{h(e)}$ and $\mu_{h(e)}$ [See Fig. \ref{Eq. S1} (a,b)].
\begin{figure}[h]
\begin{center}
\includegraphics[width=0.8\hsize]{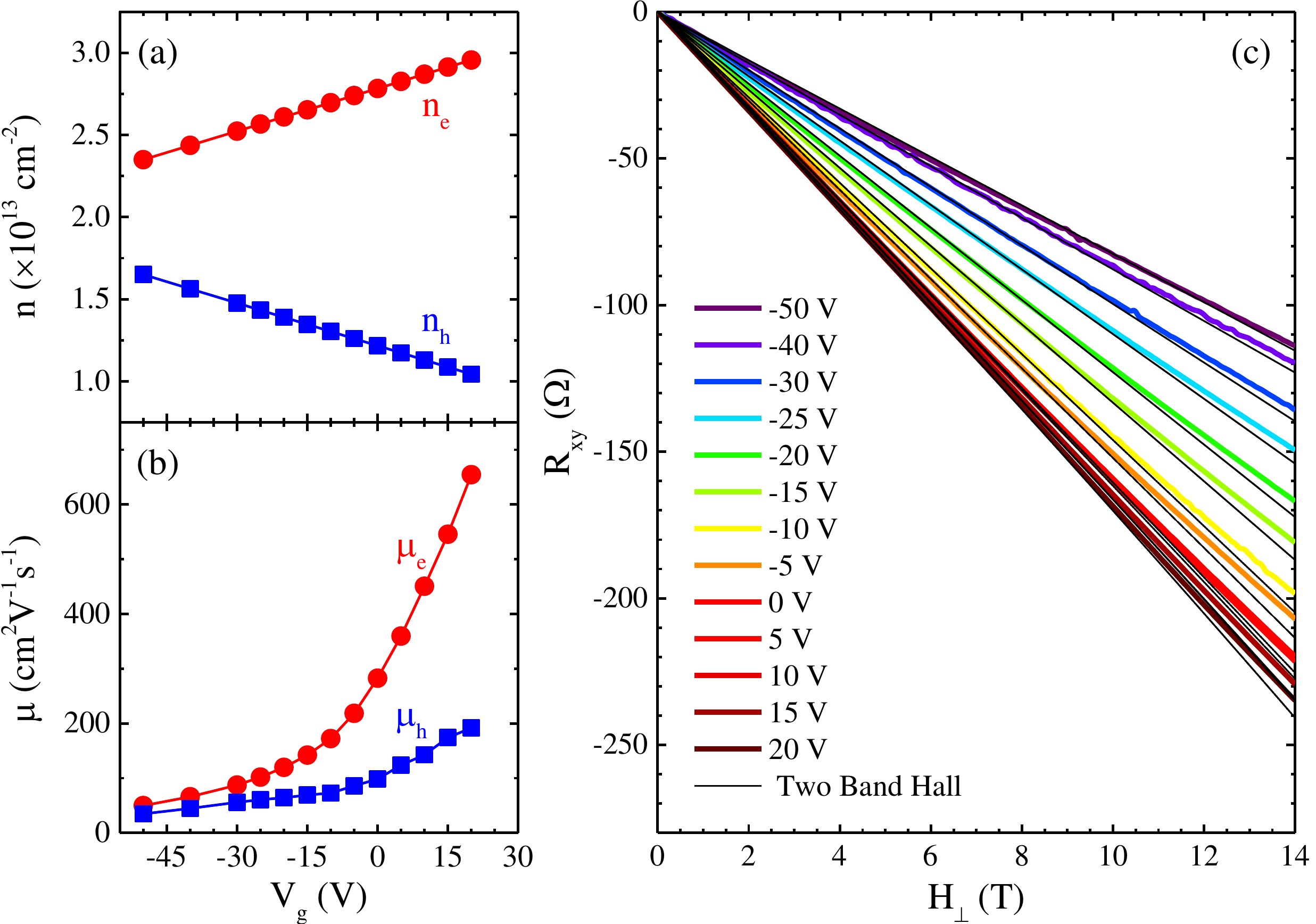}
\caption{The gate dependance of carrier density (a) and mobility (b) for electron and hole band extracted from two band model. (c) The Hall resistance ($R_{xy}$) as a function of magnetic field for different \Vg~measured at 5 K. The solid curves are plotted using the extracted $n_{h(e)}$ and $\mu_{h(e)}$ in Eq. \ref{Eq. S2}.}
\label{Fig-HallTwoBand}
\end{center}
\end{figure}

\pagebreak

Using the extracted $n_{h(e)}$ and $\mu_{h(e)}$, we have calculated the Hall resistance for high magnetic fields by the expression:
\begin{equation}\label{Eq. S2}
{R_{xy}} = \frac{1}{e}\frac{{({n_h}\mu _h^2 - {n_e}\mu _e^2) + \mu _h^2\mu _e^2({n_h} - {n_e}){H^2}}}{{{{({n_h}{\mu _h} + {n_e}{\mu _e})}^2} + \mu _h^2\mu _e^2{{({n_h} - {n_e})}^2}{H^2}}}H.
\end{equation}
The calculated Hall curves have good agreement with the measured data for lower fields [see Fig. \ref{Fig-HallTwoBand} (c)]. However, we see more deviation with increasing field. This can be due to more complicated effects such as splitting of the electron band into two spin states with an avoided band crossing due to spin-orbit interaction. Our simplified two-band description cannot capture these effects. We want to point out that, in order to perform a more accurate analysis, one needs additional measurements such as Shubnikov-de Haas oscillations, which can accurately determine the carrier density in the mobile band \cite{maniv2015}. Despite its simplicity our analysis provides the parameter regimes for the electron and hole bands and the qualitative trend of these parameters.

\par
As expected, ${n_e}$ increases with increasing \Vg~while ${n_h}$ follows the opposite trend [see Fig. \ref{Eq. S1} (a)]. However, the mobility increases with \Vg~ for both bands. The drastic rise in mobility $\mu_e$ with increasing \Vg~can be responsible for large variation in $R_S$. Since ${n_e}\mu _e^2 > {n_h}\mu _h^2$ for all \Vg, Eq. \ref{Eq. S1} reveals that the Hall slope should be negative as observed in Fig. \ref{Fig-HallTwoBand} (c). Therefore 1/$\left| {eR_H} \right|$ starts increasing with increasing \Vg~(or increasing number of electrons).

\section{Analysis of magnetoresistance data}

Figure \ref{Fig-WALFitting} presents gate dependence of $g$-factor as well as the coefficients $A$ and $C$ related to orbital magnetoresistance given by the last term in Eq. (1) of the manuscript. According to Drude model for a two-dimensional electron gas, the elastic scattering time $\tau$ is given by $\tau  = m^* /{e^2}{n_S}{R_S}$ ($m^*$ is the effective electron mass and $n_S$ is the carrier density) and the diffusion coefficient  $D$ can be expressed as:  $D = v_F^2 \tau/2$ (${v_F} = \hbar \sqrt {2\pi {n_S}} /m^*$ is the Fermi velocity). We have extracted $D$ for various $V_g$ and used it for the fitting of magnetoresistance data [Fig. 3 in the manuscript]. The extracted $g$ values are slightly higher than the typical values of 2 for a free electron system and slowly increase with increasing $V_g$. Similar gate dependence of $g$ has been observed in (100) interface previously \cite{caviglia2010}. $A$  and $C$ are related to the mobility ($\mu$) and as expected increases with $\mu$ when the gate voltage $V_g$ is increased. However, the exact dependencies of these parameters on other measurable transport parameters (like $R_S$, $n_S$ etc.) are much more complicated due to multiband charge transport.

\begin{figure}[h]
\begin{center}
\includegraphics[width=0.8\hsize]{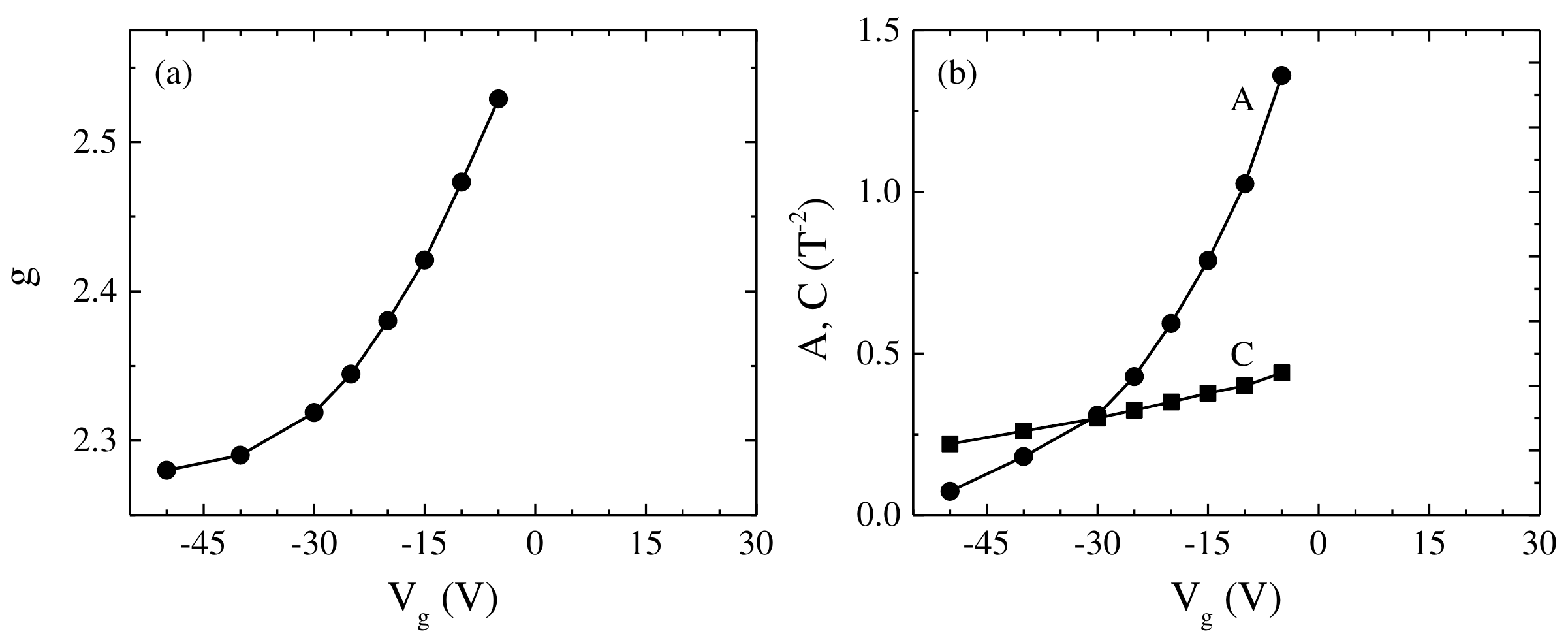}
\caption{Gate dependence of the fitting parameters $g$ (a), $A$ (b) and $C$ (b) at 1.3 K.}
\label{Fig-WALFitting}
\end{center}
\end{figure}

We can determine the spin-orbit time $\tau _{SO}$ and the inelastic time $\tau _{i}$ using the relations ${H_{i ,SO}} = \hbar /4eD{\tau _{i ,SO}}$. Therefore, we can determine $\tau _{SO}$, $\tau _{i}$, and $\tau$ using the experimental values of $R_S$, $n_S$ (or 1/$\left| {eR_H} \right|$), the fitting parameters $H_{i }$, $H_{SO}$, and a typical $m^* =$ 3$m_e$ ($m_e$ is the electronic mass). These three time scales are presented in Fig. 3(d) of the manuscript.

\bibliographystyle{apsrev}

\begin{thebibliography}{40}
\bibitem[{Rem()}]{Remark}
\bibinfo{note}{For $V_g \leq$ -40V the saturation resistance at high field is
  significantly larger than the resistance at 350mK (above $T_c$). But these
  two quantities coincide for other gate voltages. This may be related to the
  proximity to a superconductor-to-insulator transition for lower $V_g$ and
  will be discussed elsewhere.}
\end{thebibliography}

\end{widetext}

\end{document}